
\input phyzzx
\overfullrule=0pt

\def\ebnbjj{$q\bar{q} \ra t \bar t \ra be^+\nu\bar{b}q_1\bar{q}_2$}

\def\gtt{\hbox{$g\!-\!t-\!\bar t$}}

\def\nonumber{{}}
\def\frac#1#2{{#1 \over #2}}
\def\qqtt{$q\bar{q}\rightarrow t\bar{t}$}
\def\up{\uparrow}
\def\down{\downarrow}

\def\ra{\rightarrow}

\def\half{{1 \over 2}}

\def\journal#1&#2(#3)#4{
{\unskip,~\sl #1\unskip~\bf\ignorespaces #2\unskip~\rm (19#3) #4}}
\def\jour#1&#2(#3)#4{
{\unskip ~\sl #1\unskip~\bf\ignorespaces #2\unskip~\rm (19#3) #4}}
\def\Eq#1{Eq.~{#1}}
\def\fg#1{Fig.~{#1}}

\nopubblock
\vbox{
\line{\hfil  FSU-HEP-930508}
\line{\hfil  MSUHEP-93/04}
\line{\hfil May 1993}
\titlepage
\title{ Leading Electroweak Corrections to the Production of Heavy Top Quarks
at Hadron Colliders}
\author{C.~Kao}
\address{
Department of Physics \break
Florida State University \break
Tallahassee, FL 32306-3016}
\author{ G.~A. Ladinsky \ \ and\ \ C.--P. Yuan}
\address{
Department of Physics and Astronomy \break
Michigan State University \break
East Lansing, MI 48824-1116}
\abstract{
We calculate the electroweak corrections of the order
$O({m_t^2\over v^2})$ to the QCD production of $t\bar{t}$ pairs via
$q \bar q \ra t\bar t$ at hadron colliders and show that
these corrections to the total production rate are small.
This correction can be characterized as increasing the cross
section most near the threshold region, where top quark signals
are important, while the corrections become negative at higher $t\bar{t}$
energies where the top quark is a background for heavy Higgs boson searches
or investigations involving the strongly interacting longitudinal
$W$ system.
The polarization of the $t\bar t$ pair is also discussed,
including the effect that this has on proposed techniques for measuring the
the top quark mass.
}
}
\endpage
%
%
\REF\nonpert{V.S.~Fadin and V.A.~Khoze\journal JETP~Lett.&46 (87) 525;
${}\hfill$\break
\jour Sov.~J.~Nucl.~Phys.&48 (88) 309.}
\REF\dawson{
P. Nason, S. Dawson and R.K. Ellis\journal Nucl.~Phys. & B303 (88) 607;
\jour {}&B327 (89) 49;
W. Beenakker, H. Kuijf, W.L. van Neerven and J. Smith,
\jour Phys.~Rev. & D40 (89) 54;
W. Beenakker, W.L. van Neerven, R. Meng, G.A. Schuler and J. Smith,
DESY 90-064 (90);
R. Meng, G.A. Schuler, J. Smith and W.L. van Neerven,
\jour Nucl.~Phys. & B339 (90) 325.}
\REF\tpol{G.L.~Kane, G.A.~Ladinsky and C.--P.~Yuan,
\jour Phys.~Rev.&D45 (92) 124.}
\REF\sdc{Solenoidal Detector Collaboration, Technical Design of a Detector
to be operated at the Superconducting Super Collider, 1 April 1992,
preprint SDC-92-201.}
\REF\YUAN{C.--P.~Yuan and T.~C.~Yuan\journal Phys.~Rev.&D44 (91) 3603.}
\REF\LOOP{D.~Dicus and C.~Kao, LOOP, a FORTRAN program for doing
loop integrals of 1, 2, 3 and 4 point functions with momenta in the numerator,
unpublished, (1991).}
\REF\TINI{G.~Passarino and M.~Veltman\journal Nucl.~Phys.&B160 (79) 151.}
\REF\rscreen{M. Veltman\journal Acta~Phys.~Pol.&B8 (77) 475.}
\REF\tung{J.G.~Morfin and W.K.~Tung, \jour Z.~Phys.&C52 (91) 13.}
\REF\wwww{J.~Bagger, V.~Barger, K.~Cheung, J.~Gunion, T.~Han, G.A.~Ladinsky,
R.~Rosenfeld and C.--P.~Yuan, unpublished.}
\REF\kpr{G.L. Kane, J. Pumplin and W. Repko\journal Phys.~Rev.~Lett.&{41}
(78) 1689;
A. Devoto, G.L. Kane, J. Pumplin and
W. Repko\journal Phys.~Rev.~Lett. &{43} (79) 1062;
\jour Phys.~Rev.~Lett. &{43} (79) 1540;
A. Devoto, G.L. Kane, J. Pumplin and
W. Repko\journal Phys.~Lett. & {90B} (80) 436.}
\REF\density{G.A. Ladinsky\journal Phys.Rev.&D46 (92) 2922.}
\REF\gfnal{G.A.~Ladinsky, \jour Phys.Rev.&D46 (92) 3789.}
\REF\rnonpert{M.J.~Strassler and M.E.~Peskin, \jour Phys.~Rev.&D43 (91) 1991.}
\REF\lhcconf{F.~Berends et al., in the report of the Top Physics Working
Group from the {\it Proceedings of the Large Hadron
Collider Workshop}, 4-9 October 1990, Aachen, ed. G.~Jarlskog and D.~Rein,
CERN publication CERN 90-10, pg.~310, Fig.~7/8.}
\REF\ranother{A.~Stange and S.~Willenbrock, Fermilab preprint
FERMILAB-PUB-93-027-T, Feb. 1993.}
%
%
\FIG\fone{Diagrams contributing to (a)~the top quark self energy and
wave function renormalization and (b)~the electroweak corrections to the
$gt\bar{t}$ vertex.}
\FIG\ftwo{Taking $m_t=180\,$GeV, we plot the distribution
$d\sigma /ds$ vs. $s$ from the Born level QCD subprocess \qqtt\ in
(a) proton--antiproton collisions at Tevatron energies of $1.8\,$TeV,
(b) proton--proton collisions at an LHC energy of $16\,$TeV and
(c) proton--proton collisions at an SSC energy of $40\,$TeV.}
\FIG\ffive{These figures (a)--(c) are the same as \fg{\ftwo},
except that $m_t=140\,$GeV.}
\FIG\fthree{Taking $m_t=180\,$GeV and $m_H=100\,$GeV, we plot the $K$--factor
of Eqs.{\kfnal} and {\kssc} against the subprocess center--of--mass energy
$M(t\bar{t})$ considering\break
(a) $t$ helicity states with an unpolarized $\bar{t}$
(R indicates right--handed $t$, L indicates a left--handed $t$);
(b) $t$ and $\bar{t}$ helicity states
(RL indicates right--handed $t$, left--handed $\bar{t}$, etc.)
.}
\FIG\fsix{These figures (a)--(b) are the same as \fg{\fthree},
except that $m_t=140\,$GeV.}
\FIG\ffour{These figures (a)--(b) are the same as \fg{\fthree},
except that $m_H=1\,$TeV.}
\FIG\fnew{Taking $m_t=180\,$GeV and $\sqrt{s}=500\,$GeV, we plot the single
particle asymmetry $P_\perp$ as described by Eq.~(3.7) when top quark spin is
perpendicular to the scatter plane against $\cos\theta$.  The two curves
represent the asymmetry for two different values of the Higgs boson mass.}
\FIG\fseven{These two curves represent the distribution $d\sigma/dM(eb)$ vs.
$M(eb)$ at the Tevatron for right--handed and left--handed top quark
helicities using $m_t=140\,$GeV and $m_b=0$.
Kinematic constraints in the lab frame
on the rapidity ($|\eta|<2.5$) and transverse momentum ($p_T>20\,$GeV)
were imposed on the $e^+$ and $b$ quark.}
\FIG\fnine{This plot shows the variation of the $K$ factor with the mass of the
$t\bar t$ pair ($M(t\bar{t})$) using $m_t=150,\ 200,\ 250\,$GeV and
$m_H=100\,$GeV in the unpolarized production rates.}
%
%
%
\chapter{Introduction}

The mass of the top quark is one of the yet-to-be-measured parameters in the
Standard Model (SM).
To test the SM and probe new physics, we need to
know the mass ($m_t$) of the top quark well, since
at present, we still know neither the mechanism
of electroweak symmetry breaking nor the mechanism for the generation of
fermion masses.
For instance, one might be able to
determine the mass of the SM Higgs boson from the precision test of
electroweak radiative corrections if $m_t$ is known.
{}From another perspective,
at the SSC (Superconducting Super Collider) and the LHC
(Large Hadron Collider) the top quark production rate is large enough
that it can potentially become a serious background when searching for
new physics.
Therefore, we need to know the production rate of the top quark as precisely
as possible.  Because the decay products of
the top quark can mimic the signature of signal events, such as those
involved with Higgs boson searches, it is also
useful to know the polarization of the top quark which in turn
controls the kinematics of the particles created by the top quark decays.

In this paper, we present the results for the {\it leading}
electroweak corrections of the order $O({m_t^2\over v^2})$ to the
production of $t\bar{t}$ pairs as computed for the QCD subprocess
$q\bar{q}\to t\bar{t}$.
We find that the corrections to
the total rate are small with a few percent increase in the cross
section near the threshold region for a light (around $100\,$GeV) Higgs
boson and about the same percent decrease in cross section at high
subprocess energies.  These corrections can be taken most seriously
for higher top quark masses, where not only is the $O({m_t^2\over v^2})$
approximation most valid, but also the nonperturbative effects
which can modify the threshold behavior become less significant since
the faster decay time for the top quark prevents bound states from
forming.\refmark{\nonpert}

The higher order corrections were applied using helicity amplitude
techniques in the computation of $K$--factors for both polarized and
unpolarized final states.  Though the QCD corrections to the \qqtt\
production rates\refmark{\dawson} are larger than the leading electroweak
results presented here, the parity violation manifest in the electroweak
interactions produces effects unobtainable by theories like QCD that
maintain parity ($P$) and charge conjugation ($C$) symmetries
separately.  These effects are realized, for example, in the
difference between the $K$--factors describing the higher order
corrections for final state polarizations related by parity
transformations.
Such differences are often largest as the invariant mass of
the $t\bar{t}$ pair gets large, making the polarization effects
most relevant when considering the backgrounds to signal events
like those discussed for probing the electroweak
symmetry breaking mechanism in the TeV region.

The effect that polarization has on the observed kinematics of the
\qqtt\ events is reflected in the decay products of the top
quarks.\refmark\tpol Consequently, there may be a change in the
efficiency of experimental cuts used to remove the top quark
background or to observe the top quark signal depending upon the spin
asymmetries in top quark production.  For example, the charged lepton
produced from the decay of a top quark with a right--handed helicity
via $t\to bl^+\nu$ will preferentially receive a greater boost along
the direction of motion of the top quark than the charged lepton would
from top quark with a left--handed helicity.  An enhancement of top quarks with
a right--handed (left--handed) polarization will then produce charged
leptons with more (less) energy in the laboratory.  Realizing this, it
is plausible that a fixed cut on lepton energies determined from
leading order top quark production may automatically produce a
different efficiency in removing background or collecting signals than
would be obtained from the production rate that contained electroweak
corrections.  Perhaps even more pertinent to present interests is the
effect that the polarization asymmetry has on the distribution
of $M(eb)$, the mass of the charged lepton and bottom quark system,
in $t\to e^+b\nu$, since it is through a related mass distribution
(invariant mass of the $e^+$ and $\mu^-$, $M(e\mu)$, where the muon comes from
fragmentation of the bottom quark)
that the best techniques for determining the top quark mass are
derived.\refmark{\sdc}

The remainder of this paper is organized as follows. In section 2, we
present the analytical results of our calculations in terms
of form factors. Using these form factors, we then give the numerical
results on the production rate and the degree of
polarization of the top quark in section 3.
In section~4 we examine the polarization effects on the $M(eb)$ distribution
and discuss how this relates to measurements of the top quark mass.
Section 5 contains our conclusion.

\chapter{The Loop Corrections}

The effective theory considered in this paper is obtained
by taking the limit of the electroweak coupling $g \to 0$
after replacing the mass of the $W^+$ boson ($M_W$) with $gv/2$
in the SM Lagrangian.
The neutral and charged Goldstone bosons that remain
($\phi^0$ and $\phi^\pm$) are massless.
The parameter $v \approx 246 $ GeV characterizes
the scale of the electroweak symmetry breaking and corresponds to the
vacuum expectation value (VEV) of the Higgs field in the SM.

We show that the form factors are infrared--safe, and it
is not necessary to include real diagrams with an additional
$\phi^0$ or $\phi^\pm$ associated with the
$t \bar t$ production in our calculations.
The Landau gauge has been chosen to evaluate the loop diagrams, and
the ultraviolet divergences are regularized by dimensional regularization
with the regulator $\Delta \equiv 2/(4-N) -\gamma_E +ln(4\pi)$, where $N$ is
the spacetime dimension and $\gamma_E$ is the Euler constant.

\section{Wave Function Renormalization}

There are three diagrams, as shown in \fg{\fone}(a),
contributing to the self energy of the top quark and its wave function
renormalization.
The wave function renormalization constant $Z_t$ can be written as
$$\eqalign{
Z_t     & =  1 +\frac{1}{16\pi^2} \frac{m^2}{v^2}
             [ \delta Z_t^V -\delta Z_t^A \gamma_5 ].
\cr}
\eqn\ztmod
$$
Hereafter, we use $m$ and $m_t$ interchangeable.
Employing the on-shell renormalization scheme, we obtain \refmark{\YUAN}
$$\eqalign{
\delta Z_t^V & =
- \frac{1}{2}[ 3 \Delta -3 ln(\frac{m^2}{\mu^2}) \nonumber \cr
             &    +1 -2I(r) +2J(r) +4L(r) +i\pi ] , \nonumber \cr
\delta Z_t^A & =
 \frac{1}{2} [ \Delta -ln(\frac{m^2}{\mu^2}) +2 +i\pi ] ,
\cr}
\eqn\ztvztamod
$$
where $\mu$ is the 't Hooft mass parameter, $r = m_H^2/m^2$, and
 $m_H$  is the mass of the Higgs boson.
The integrals $I(r)$, $J(r)$ and $L(r)$ are defined as
$$\eqalign{
I(r) & =  \int_0^1 ln [ x^2 +r(1-x)-i\epsilon] dx , \nonumber \cr
J(r) & =  \int_0^1 x ln [ x^2 +r(1-x)-i\epsilon] dx , \nonumber \cr
L(r) & =  \int_0^1 \frac{x(1-x^2)}{ (1-x)^2 +rx-i\epsilon} dx.
\cr}
\eqn\ijl
$$

\section{Vertex Corrections}

The $gt\bar{t}$ vertex can be expressed as
$$
i g_s  \bar{u}(p) T^a \Gamma_\mu v(q),
\eqn\vertex
$$
where $g_s$ is the strong coupling and the $T^a$ are the $SU(3)$ matrices
with \hbox{$Tr(T^aT^b)={1 \over 2} \delta^{ab}$}.  The
$u(p)$ and $v(q)$ are the Dirac spinors of the $t$ and $\bar{t}$ with momenta
$p$ and $q$, respectively.

The tree level vertex function is $\Gamma_\mu^{tree} = \gamma_\mu$.
At the 1-loop level, as shown in \fg{\fone}(b), the vertex function can be
written as
$$\eqalign{
\Gamma_\mu^{loop} & =  \frac{1}{16\pi^2} \frac{m^2}{v^2}
                        \bar{u}(p) \Lambda_\mu v(q) , \nonumber \cr
\Lambda_\mu       & =  \gamma_\mu (A -B\gamma_5) \nonumber \cr
                  &    +\frac{1}{2}(p_\mu-q_\mu)(C -D \gamma_5)
                        \nonumber \cr
                  &    +\frac{1}{2}(p_\mu+q_\mu)(E -F \gamma_5) ,
\cr}
\eqn\vtxfun
$$
where
$$\eqalign{
A & =   \frac{3}{2}[ \Delta -ln(\frac{m^2}{\mu^2}) +1  ]
            +\frac{1}{4}( \frac{1}{\beta^2} -3 )ln(\frac{s}{m^2})
            -\beta ln(\frac{1+\beta}{1-\beta}) \nonumber \cr
      &    +\frac{m_H^2}{s\beta^2}
             [ -I(r) -2 +\beta ln(\frac{1+\beta}{1-\beta}) ] \nonumber \cr
      &    -( \frac{m_H^4}{s\beta^2} +4m^2 )C_0^H
            -\frac{s}{16}( \frac{1}{\beta^2} +2 -
3\beta^2 )C_0^{\phi^+}\nonumber \cr
      &    +i \pi( -\frac{m_H^2}{s\beta} +\frac{1}{2} +\beta ) , \nonumber \cr
B & =  -\frac{1}{2}[ \Delta -ln(\frac{m^2}{\mu^2}) +1 ]
           +\frac{1}{4}( \frac{3}{\beta^2} -1 )ln(\frac{s}{m^2}) \nonumber \cr
      &   +\frac{s}{16}(-\frac{3}{\beta^2} +2 +\beta^2 ) C_0^{\phi^+}
           -i (\frac{\pi}{2})  , \nonumber \cr
C & =  \frac{m}{s\beta^2} \{
             ( \frac{3}{\beta^2} -1) ln(\frac{s}{m^2})
            -18 -8I(r) +4 \beta ln(\frac{1+\beta}{1-\beta}) \nonumber \cr
      &    +(\frac{12 m_H^2}{s\beta^2})[ -I(r) -2
            +\beta ln(\frac{1+\beta}{1-\beta}) ]
            +2 r [ I(r) -ln(r) +1 ]\nonumber \cr
      &    -12 m_H^2 ( \frac{m_H^2}{s\beta^2} +1 )C_0^H
            -\frac{s}{4}( \frac{3}{\beta^2} -2 -\beta^2 ) C_0^{\phi^+}
            -4i\pi (\frac{3m_H^2}{s\beta} +\beta) \} , \nonumber \cr
F & =   \frac{m}{s} [(1- \frac{3}{\beta^2} ) ln(\frac{s}{m^2}) -2
            +\frac{s}{4} ( \frac{3}{\beta^2}
                -2 -\beta^2 ) C_0^{\phi+} ]  , \nonumber \cr
D & =  E = 0,
\cr}
\eqn\frmfac
$$
and
$$\eqalign{
C_0^H     & =  C_0(m^2,m^2,s,m^2,m_H^2,m^2) , \nonumber \cr
C_0^{\phi^+} & =  C_0(m^2,m^2,s,0,0,0),
\cr}
\eqn\scaints
$$
given that $s = (p+q)^2$ is the squared of the $t \bar t$
center-of-mass energy and $\beta = \sqrt{1- 4 m^2/s} $.  The form
factor $D$ is zero because this theory is CP invariant.  We note from
Eq.~(2.9) that
vector current conservation demands that $\delta Z^A_t +B=-s F /4m$.

The 3-point function $C_0$ is defined as
$$\eqalign{
C_0(p_1^2,p_2^2,p_5^2,m_1^1,m_2^2,m_3^2) &  \equiv  \nonumber \cr
   \frac{1}{i \pi^2} & \int d^N q \frac{1}{[q^2 -m_1^2][(q+p_1)^2 -m_2^2]
[(q+p_1+p_2)^2 -m_3^2]},
\cr}
\eqn\theint
$$
where $p_5 = p_1 +p_2$.
The loop integrals in the form factors have been evaluated with the code LOOP.
\refmark{\LOOP,\TINI}
For simplicity, the mass of the bottom quark ($m_b$)
is taken to be zero.\footnote{\dagger}{We checked that the difference between
using $m_b = 5 $ GeV and zero is less than $0.1\%$ in the numerical results of
the form factors. }

The renormalized vertex function becomes
$$\eqalign{
\Gamma_\mu^R
             & =  \gamma_\mu \nonumber \cr
             &    +\frac{1}{16\pi^2} \frac{m^2}{v^2} \{
                   \gamma_\mu [ ( \delta Z^V_t +A  )
                   -( \delta Z^A_t +B ) \gamma_5] \nonumber \cr
             &    +\frac{1}{2}(p_\mu-q_\mu)(C -D \gamma_5)
                   +\frac{1}{2}(p_\mu+q_\mu)(E -F \gamma_5)  \}.
\cr}
\eqn\renvfun
$$
It can be seen explicitly that the terms with the regulator, $\Delta$,
and the mass parameter, $\mu$, cancel exactly among themselves, therefore
the renormalized vertex function is free of ultraviolet divergence and
independent of the $\mu$ parameter as expected.
In addition, it has been checked that $\Gamma_\mu^R$ is free of
infrared divergence.

When the Higgs mass is very large, $i.$ $e.$, $m_H^2 \gg s > m^2$,
$$\eqalign{
rI(r)  & \to  -{1 \over 2} + rln(r) +ln(r) -r ,\nonumber \cr
J(r)  & \to  -\frac{3}{4} +\frac{1}{2}ln(r) ,\nonumber \cr
L(r)  & \to  0 ,\nonumber \cr
C_0^{\phi^+} & \to  0 ,\nonumber \cr
C_0^H & \to
-\frac{1}{m_H^2}[ ln(r) +1 -\beta ln( \frac{1+\beta}{1-\beta} ) +i \pi \beta]
\nonumber
+\frac{1}{m_H^4} ln(r) ( \frac{s}{2} -3m^2 ) .
\cr}
\eqn\lghiggs
$$
It is straightforward to check that the dependence of the
renormalized vertex function
$\Gamma_\mu^R$ on Higgs mass vanishes as $m_H\to\infty$.
(Recall that $r=m_H^2/m^2$.)
Therefore, $m_H$ decouples in this case for the heavy Higgs mass limit.
This is in contrast to the usual one loop SM electroweak corrections
which grow like $ln(m_H)$ in the heavy
Higgs mass limit.\refmark\rscreen\

\chapter{Numerical Results}

\section{General}

It has been demonstrated that the modifications of the \gtt\ vertex due to
the electroweak corrections given by \Eq{\renvfun} appear as finite
modifications to the form factors.
In particular, the values of $E$ and $F$ are irrelevant for the \qqtt\ process
considered, as these terms will
vanish when the momentum factor in front of them couples to the
annihilation vertex for the massless quarks in the initial state.
With the corrections written in the
manner of \Eq{\renvfun}, it becomes possible to use helicity amplitude
techniques for
computing the loop effects in $t\bar{t}$ production.\refmark{\tpol}
In particular,
this allows us to preserve the polarization information.

In \fg{\ftwo} and \fg{\ffive} we present the variation of the hadron
level distribution with the subprocess center--of--mass energy
for the Fermilab Tevatron, the LHC, and the SSC for $m_t=180\,$GeV and
$m_t=140\,$GeV, respectively. We use the parton distribution functions
of Morfin and Tung\refmark{\tung} (set SL) using a scale of $Q=\sqrt{s}$.
This same scale, which is the center--of--mass energy of the $t\bar t$ pair,
is also used in evaluating the strong coupling constant.
We note that at the
SSC and the LHC, the dominant production mechanism for the $t \bar t$
pairs is $gg \to t \bar t$, yet in the very high invariant mass region
(about or above 1 TeV) $q \bar q$ fusion can be important as
a background to the study of the electroweak symmetry
breaking.\refmark\wwww\

\subsection{Definition of $K$--Factor}

Denoting the higher order cross section which includes the leading electroweak
corrections up to $O(\alpha_s^2{m_t^2/v^2})$ at the parton level as
$\hat{\sigma}^{H.O.}$, we define the $K$--factor,
$$
\hat{K}(s,m_t,m_H)\equiv
{d\hat{\sigma}^{H.O.}\over ds}/ {d\hat{\sigma}^{Born}\over ds},
\eqn\kfacdef
$$
which quantifies the leading electroweak corrections
to the parton cross section at the Born level, $\hat{\sigma}^{Born}$.
For processes like the $s$--channel \qqtt\ considered here, this
$K$--factor is valid also at the hadron level for all system rapidities,
$$
y=\half\ln {x_a \over x_b},
\eqn\rapdef
$$
given our choice of scale,
where $x_a,x_b$ are the fractions of momenta that the $q$ and $\bar{q}$
take from their parent hadrons.

The hadron level $K$--factor is defined as
$$
K(s,m_t,m_H)\equiv
{d{\sigma}^{H.O.}\over ds}/ {d{\sigma}^{Born}\over ds}.
\eqn\kfachad
$$
\Eq{\kfachad} differs from \Eq{\kfacdef} in that the ratio is with regards
to the hadronic differential cross sections, which are simply the parton
differential cross sections convoluted with the parton distribution functions,
$$
\sigma=\sum_{a,b}\int dx_a dx_b f_{a/A}(x_a,Q)f_{b/B}(x_b,Q)
d\hat{\sigma}(a+b\to t+\bar{t}),
\eqn\hadxsec
$$
where $f_{a/A}(x_a,Q)$ provides the density for partons of flavor $a$
carrying momentum fraction $x_a$ of the total momentum of hadron $A$.
As with the strong coupling, the scale $Q$ in the parton densities
has been set to $\sqrt{s}$.
After converting the $dx_adx_b$ integrals over the parton momentum fractions
to $dyds/S$ integrals in \Eq{\hadxsec}, where
$S=s/x_ax_b$ is the center--of--mass energy of the hadron--hadron system,
the parton cross section factors
outside the $dy$ integration because the parton level cross section depends
only on $s$ and masses. (It is implied that the integration over top polar
angle has been performed.)
Computing the integral over $dy$ yields a cancellation of the contribution from
the parton distribution functions between the numerator and
the denominator in \Eq{\kfachad},
such that the parton level $K$--factor of \Eq{\kfacdef}
is the same as the analogous hadron level $K$--factor of \Eq{\kfachad}.

This is independent of whether the initial hadrons are protons or antiprotons.
So, the parton level $K$-factors presented may be considered as the
hadron level $K$--factors at the Fermilab Tevatron,
$$
\hat{K}(s,m_t,m_H)=K_{FNAL}(s,m_t,m_H)\equiv
{d{\sigma}^{H.O.}_{FNAL}\over ds}/ {d{\sigma}^{Born}_{FNAL}\over ds},
\eqn\kfnal
$$
and at the SSC and LHC,
$$
\hat{K}(s,m_t,m_H)=K_{SSC/LHC}(s,m_t,m_H)\equiv
{d{\sigma}^{H.O.}_{SSC/LHC}\over ds}/
			{d{\sigma}^{Born}_{SSC/LHC}\over ds},
\eqn\kssc
$$
where $\sigma^{H.O.}$ and $\sigma^{Born}$ respectively represent the
hadron level cross sections for the production of $t\bar{t}$ pairs
through quark--antiquark annihilation at
the one loop level and at the tree level.
Note that \Eq{\kfnal} and \Eq{\kssc}
are true provided that no kinematic cuts are applied
to the $t$ or $\bar t$.

One of the advantages of computing the higher order correction through
modification of the form factors
is that we can conveniently implement the corrections at the
amplitude level and examine the consequent changes in the production
of polarized top quarks.  The higher order effects vary somewhat when
we compare the production of unpolarized $t\bar{t}$ pairs to the
production of polarized final states.
The results for polarized top quarks are given in Figs.~\fthree--\ffour.

\subsection{Magnitude of Results}

In general the higher order electroweak effects in \qqtt\ yield only a
small correction of a few percent to the cross section.  In \fg{\fthree}
and \fg{\fsix} the $K$--factors that describe these corrections
are greater than unity near the threshold region for a light Higgs
boson, reaching magnitudes around 1.08 (1.03) for
$m_t=180\ (140)\,$GeV and $m_H=100\,$GeV;
a drop in value occurs as
we go to subprocess energies of $3\,$TeV providing a negative correction
to the Born level rates of no greater than a ten percent reduction.
For the production of $t\bar{t}$ pairs in the threshold region
given a heavy Higgs boson (see \fg{\ffour}), we find a small
decrease in rate yielding $K$--factors just under unity.  Despite the
large size of the $K$--factor for the lighter Higgs boson when $s$ is
extremely close to the mass threshold of producing the $t \bar t$
pair, the effect on the total cross section is small because of the
suppression of the phase space indicated in Figs.~\ftwo\ and \ffive.
At subprocess energies far from threshold, the event rate for top
quark production is much smaller; nevertheless, it is useful to know
that the electroweak corrections cause a decrease in the event rate for
large invariant masses, $M(t\bar{t})$, of the $t\bar{t}$ pairs
because it is in the high invariant mass region at the SSC/LHC
that the top quark is
a background to signals needed to study the electroweak
symmetry breaking sector (given that no light Higgs
boson is found).\refmark{\wwww}

The $K$--factor has a dependence on both the mass of the top quark and
the mass of the Higgs boson, both unknown quantities at this time.
The general outcome of an increase in top quark mass from $140\,$GeV
to $180\,$GeV is that $|1-K|$ is slightly larger for the heavier quark
near threshold and at high $M(t\bar{t})$.
If we fix
$m_H$ at either $100\,$GeV or $1\,$TeV while changing the top quark
mass from $140\,$GeV to $180\,$GeV, we find the $K$--factor deviation
from unity is about a factor of two or three greater for the lighter Higgs
boson mass.  From the perspective of fixed $m_t$, we also see the
variation in the $K$--factor between $m_H=100\,$GeV and
$m_H=1\,$TeV is larger for the larger top quark mass.
As mentioned previously, in the limit that mass of the Higgs boson is taken to
infinity, the renormalized vertex function $\Gamma_\mu^R$ in
\Eq{\renvfun} loses its dependence on $m_H$.

\section{New Effects That Did Not Appear at the Tree Level}

Though the effect of the electroweak corrections are small when
compared to the total cross section, there are conditions where the
form factor modifications can yield a relatively significant change in
a particular production mode.  In particular, because the
form factors become complex, there are polarization asymmetries which
develop nonzero values in contradistinction to their Born level
counterparts.  Such is the case when considering
inclusive cross sections for top quark production where the polarization
of the observed top quark is transverse to the scatter plane.

As was discussed in Ref.~{\tpol}, the Born level amplitudes for
$q\bar{q}\to t\bar{t}$ are all real.
For this reason, a single particle asymmetry is zero when
considering the transverse polarization perpendicular to the scatter
plane.\refmark{\kpr}  In computing the higher order corrections, however,
an imaginary portion is generated in the form factors that takes the
polarization perpendicular to the scatter plane to nonzero values.
These nonzero values can in principle be used to test for CP
violation effects.\refmark\tpol

The radiative corrections computed here make the helicity amplitudes
complex in value.  This produces a nonzero value for the polarization
of single top quark spins directed perpendicular to the scatter plane,
$P_\perp$.  Define $P_\perp$ by
$$
{d\hat{\sigma}^{\up} /d\cos\theta -d\hat{\sigma}^{\down}/d\cos\theta
\over
d\hat{\sigma}^{\up}/d\cos\theta +d\hat{\sigma}^{\down}/d\cos\theta},
\eqn\perpdef
$$
where $\hat{\sigma}^{\up}$ ($\hat{\sigma}^{\down}$) describes the parton level
cross section when the transverse spin of the top quark is pointing ``up''
(``down'') with respect to the scatter plane.  (Think of ``up'' as the
$+Y$ direction given that the \qqtt\ hard scattering occurs in the
$X$--$Z$ plane with the initial quark moving in the $+Z$ direction
and the top quark carrying a positive value for its $X$ component
of momentum.)
Though $P_\perp$ carries nonzero values,
numerical results indicate that the polarization
for single top quark spins directed perpendicular to the scatter plane
are small, yielding $P_\perp$ values not much larger than $10^{-3}$
for $m_t=180\,$GeV.
In \fg{\fnew}, we present curves of $P_\perp$~vs.~$\cos\theta$,
where $\theta$ is the angle the top quark subtends with the incoming
quark in the center--of--mass frame of the subprocess.
We found  an interesting
qualitative change in the plots as the Higgs boson mass increases.
For the lower Higgs boson masses (around $ 100\,$GeV) the polarization is
positive for $\cos\theta <0$ while for the higher Higgs boson masses
(around $1\,$TeV) the polarization for $\cos\theta <0$ is negative.
Such quantities are mainly relevant for proton--antiproton collisions,
where a convenient means is available for determining ``up'' and
``down'' directions with respect to the scatter plane by defining the
scatter plane with the proton and antiproton beams as opposed to the
annihilating quark and antiquark.
Although the $P_\perp$ plots exhibit this interesting feature,
it might be extremely difficult to measure this polarization because
of its small value.

Another new effect which is absent at the tree level is the double
polarization asymmetry $P_\perp(in,out)$ which is also sensitive to higher
order corrections.\refmark{\density} Specifically,
$P_\perp(in,out)$ refers to the
asymmetry produced when the top quark spin is perpendicular to the
scatter plane while the transverse spin of the top antiquark is in the
scatter plane.  This quantity, analogous to \Eq{\perpdef}, is zero at
the Born level and only achieves its nonzero value because of the
imaginary correction to the form factors.  Analogous to the single
spin asymmetry presented for $P_\perp$, $P_\perp(in,out)\sim 10^{-3}$
for $m_t=180\,$GeV.

So, though we can say we have an effect with $K=\infty$, the statistics
are too poor for any reasonable study.

\section{$K$--Factors for Spin Effects Present at the Born Level}

Because QCD is C (charge conjugation) and P (parity) invariant, the
single particle polarization of the top quark has to vanish at the
tree level for the process $q \bar q \to t \bar t$.  Nevertheless, the
top quark can have a single particle polarization if weak effects are
present in their production.  The effects of top quark polarization
for the Born level electroweak reaction $q\bar{q}\to(\gamma,Z)\to t\bar t$
were discussed in Ref.~{\gfnal}.  The
contribution of this process to the total cross section for the
production of $t\bar{t}$ pairs is small (about a percent at the
Tevatron), so any spin effects present in $q\bar{q}\to(\gamma,Z)\to t\bar t$
are diluted by the QCD
production of $t\bar t$ pairs.  Considering this larger rate for the
QCD production of $t\bar t$ pairs, similar spin effects that appear when
considering the degree of polarization due to the leading electroweak
corrections to $q\bar{q}\to g\to t\bar{t}$ at the loop level are more
significant.
The degree of polarization for a single top quark due to
leading electroweak corrections can be obtained from the curves (a)
in Figs.~\fthree--\ffour.  Typically, this effect is of the order of
a few percent for large $M(t\bar{t})$.

Besides examining single particle polarizations, there are also double
particle asymmetries in the spin dependence which can be investigated.
In the following, we consider the longitudinal and transverse spins of
the top quark and top antiquark.

\subsection{Longitudinal Spins}

When considering the longitudinal polarizations for the top quarks,
we classify the states as carrying either right--handed (R) or left--handed (L)
helicity.  In the figures and the text the correlated spin states for the
$t\bar{t}$ pairs will be labelled either RR,RL,LR,LL, where the first
letter is the top helicity and the second letter is the top antiquark
helicity.  Since the interactions described by  \Eq{\renvfun}
conserve CP, the higher order corrections make no distinction between
the RR and LL states because they are CP transforms of each other.
For all cases of $m_H$ and $m_t$ considered, the $|1-K|$ value
for the LR state is larger than that for the RL state
when the $K$--factors for both of these helicity combinations are below unity.
The reverse is true when these $K$--factors are above unity.
When $m_H$ becomes larger, the $|1-K|$ values of the $RR,LL$ spin states
become smaller, as shown in Figs.~\fthree~and~\ffour.

\subsection{Transverse Spins}

Though we do not present plots of the K--factors when both the
$t$ and $\bar t$ quarks are
polarized transverse to their direction of motion, we discuss some of the
results here.
We consider transverse spins for the top quark and antiquark to be either
perpendicular to the scatter plane or within it.  Since both quarks are
considered simultaneously, the $t,\bar{t}$ spins are further classified
as being either aligned or antialigned.

For the case where both $t,\bar{t}$ spins are perpendicular to the
scatter plane, we found the $K$--factor covering the widest range of
all our plots as we move through values of $s$.
As guided by the unpolarized results, for $m_t=180\,$GeV and $m_H=100\,$GeV
the threshold effect nears $K=1.08$,
while as we move to larger $\sqrt{s}$, the $K$--factor decreases
to about $K=0.94$.  Throughout the kinematic region considered,
the $K$--factor for when the $t$ and $\bar{t}$ spins are aligned or
antialigned remains near the $K$--factor curve for the unpolarized final state.
For the lighter top quark ($m_t=140\,$GeV), the
$K$--factor is not as large and the difference between aligned and antialigned
top spins is just as indistinct.
We also note that for the lighter top quark
the decrease in the $K$--factor near threshold is not as steep as when the
top quark was heavier.

Consider the case where the transverse spins for the two top quarks
live in the scatter plane.
For the higher $m_t$ the $K$ factor drops to
around $K=0.94$ at large $\sqrt{s}$ (around $1\,$TeV),
but no matter whether $m_t=140\,$GeV or $m_t=180\,$GeV,
the configuration where the transverse spins are aligned receives more
suppression from the high order corrections than the configuration where
the two top spins are antialigned in the TeV region.  Here the $K$ factor
doesn't stray from the unpolarized result by more than about $\pm 0.01$.

\chapter{Polarization and Top Quark Mass Measurements}

In Ref.~{\sdc} the most effective method considered for measuring
the mass of the top quark concentrated on the analysis of the invariant mass
distribution, $M(e\mu)$, which is determined from the combined momentum of the
charged $e^+$ lepton from the decay $t\to be^+\nu$ and the muon from the
fragmentation of the bottom quark.  An error of about $1.6\%$ was
estimated in the determination of the top quark mass for
$m_t=150,250\,$ GeV using a series of
kinematic cuts on the unpolarized production of top quarks.  It is
known, however, that if a polarization asymmetry were to develop in
the top quark production that the kinematics of the observed particles
would change.  If there were no kinematic cuts, this would be of
no consequence since integrating out the angular dependence
washes out
the polarization effects on this measurement; however, with kinematic cuts,
as required in reality,
a polarization asymmetry can affect the $M(e\mu)$ mass spectrum.  With
this observation it becomes necessary to investigate the effects such
an asymmetry may produce and whether it interferes with the
precision of the mass measurement.

We proceed by examining an analogous quantity, namely, the invariant mass
of the bottom quark and charged
lepton from the top decay.  Respectively denoting the momenta of the
$e^+$, $\nu$, $b$ quark, $W$--boson, and $t$ quark as
$p_e$, $p_\nu$, $p_b$, $p_W$, $p_t$, the amplitude squared for the three--body
decay $t\to bW^+\to be^+\nu$ is given by
$$
|M|^2=
{64G_F^2m_W^4\over (p_W^2-m_W^2)^2+m_W^2\Gamma_W^2}
(p_\nu\cdot p_b)[(p_e\cdot p_t)-m_t(p_e\cdot s)],
\eqn\ampsq
$$
where $G_F$ is the Fermi coupling constant and
$s$ describes the polarization of the top quark.\refmark{\gfnal}
The neutrino and positron have been taken as massless and the masses
of the top quark, $W$--boson and bottom quark are given by $m_t$, $m_W$, $m_b$.
With the conventions chosen, the top decay rate is given by
$$
d\Gamma_t={1\over 2 m_t} |M|^2 d\Phi_3,
\eqn\gendk
$$
where the three--body phase space is
$$
d\Phi_3=
{1\over 32m_t^2(2\pi)^5}dM(eb)^2 dm_W^2 d\Omega_b d\phi_\nu^*
\Theta[-G(M(eb)^2,p_W^2,m_t^2,0,m_b^2,0)]
\eqn\lips
$$
with
$$\eqalign{
G(M(eb)^2,p_W^2, & m_t^2,0,m_b^2,0)= \cr
&2p_W^2(m_b^2M(eb)^2-m(eb)^4-M(eb)^2p_W^2-m_b^2m_t^2+M(eb)^2m_t^2).
\cr}
\eqn\thearg
$$
We choose to perform the calculation in the rest frame of the top decay.
The angular dependence in the phase space factor of \Eq{\lips} is
comprised of the differential for
the solid angle of the  bottom quark, $d\Omega_b=d\cos\theta_bd\phi_b$, and
$d\phi_\nu^*$, which is the azimuthal angle of the neutrino
measured from the coordinate system that is rotated such that the bottom quark
momentum defines the $z$--axis.

For the decay of unpolarized top quarks, \Eq{\ampsq} indicates that there
is no angular dependence to the $M(eb)$ distribution.  The phase space
integration may be easily performed in the narrow width
approximation yielding the unpolarized decay distribution,
$$
{d\Gamma_t\over dM(eb)^2}=
{G_F^2\over 16 \pi^2\Gamma_W}{\left( m_W\over m_t\right)^3}
(m_t^2-m_W^2-M(eb)^2)(m_W^2-m_b^2+M(eb)^2).
\eqn\twidth
$$
It is also clear from \Eq{\ampsq} that for polarized top quark decay
there is a spin component that
contributes to the behavior of the $M(eb)$ distribution, and this
term does have an angular dependence.
To demonstrate the effect of the spin--dependent term in \Eq{\ampsq},
we plot the $M(eb)$ distribution for \ebnbjj\
at the Tevatron in \fg{\fseven},
separating the contributions for left--handed and right--handed
helicities of the top quark.  These curves were created
for top quarks of mass $140\,$GeV
by restricting the rapidities of the $e^+$ and $b$ quark to within 2.5
in magnitude and insisting that the transverse momentum for each of these two
particles be greater than $20\,$GeV for top quark production via
$q\bar{q}\to g\to t\bar{t}$ at the Tevatron.
(In our result we also impose the same rapidity and transverse momeuntum cuts
for the $\bar{b},q_1,\bar{q}_2$.)
A difference in the two curves for pure helicity states,
created purely by the kinematic constraints, is realized mainly in the
low $M(eb)$ region.
To understand how this affects the $M(e\mu)$ mass distribution and the
measurement of the top quark mass, one has to convolute our results
with the hadronization of the bottom quark to produce the muon.
This is beyond the scope of this paper.

\chapter{Conclusion}

We have computed the leading electroweak corrections,
of the order $O({m_t^2\over v^2})$, for \qqtt\ and found the
corrections to provide an increase in the total
cross section of no more than a few percent, which is
smaller than the typical uncertainty in the prediction
of the top quark event rate in the usual QCD processes.

A decrease appeared in $d\hat{\sigma}/d{s}$ of no
greater than ten percent compared to the lowest order result for
subprocess center--of--mass energies from around $1\,$TeV.
The perturbative results indicate an increase in the $K$--factor
just under $10\%$ near the threshold region for the $m_t=180\,$GeV and
$m_H=100\,$GeV values considered here, but this does not
include any relevant nonperturbative physics.\refmark\rnonpert\
Small transverse polarizations
were obtained from the imaginary contributions to the form factors
(generated by the loop corrections), as we found that the
polarization when considering solely the spin of the top quark perpendicular
to the scatter plane was about $10^{-3}$.

Given that our results are valid for high top quark masses and large
center--of--mass energies, we find our $K$--factors in disparity with
the results presented for the LHC in Ref.~{\lhcconf}, where the full
electroweak corrections were shown to produce a large reduction around
$40\%$ in the Born level rate for \qqtt\ at large $\sqrt{s}$ values.
In \fg{\fnine} the $K$ factors are shown for
$m_t=150,\ 200,\ 250\,$GeV using $m_H=100\,$GeV.  As $\sqrt{s}$ enters
the TeV regime, the variation in the $K$ factor becomes very flat,
never indicating a change in the Born level rate of more than $20\%$.

The parity violation
due to the electroweak couplings appears in the $K$--factors for the
production of polarized $t\bar{t}$ pairs, where for $m_H=100\,$GeV
the RL states generally received a larger $K$--factor enhancement
near threshold than that for the production of LR
states, while the $K$--factor suppression the LR states received in the
TeV region was greater than the suppression for the production of RL states.
With $m_H=1\,$TeV we saw that all helicity combinations for the final state
were suppressed, though the $K$--factor was close to unity near threshold.
We also showed that the invariant mass spectrum of $M(e\mu)$ depends on the
polarization of the top quark.  To ensure that $M(e\mu)$ is a good variable
for measuring the mass of the top quark, one has to take the effect of the
top quark polarization into consideration when performing the
analysis.

While this paper was being completed, we became aware of similar research
by Stange and Willenbrock\refmark{\ranother}
which overlaps in part with our work.
Our results agree with theirs in the total event rate.

\vbox{
\vskip 0.5cm
{\bf Acknowledgements}
\vskip 0.5cm
We would like to thank Jiang Liu for discussions.
C.K. was supported in part by the U.~S. Department of Energy under
contract number DE-FG05-87ER40319.
The work of G.A.L. and C.P.Y. was supported in part by TNRLC grant \#RGFY9240.
}
%
%
\vfill\eject
\refout
\vfill\eject
\figout
\vfill\eject
\end